\documentclass[10pt,preprint2]{aastex}

\DeclareMathAlphabet{\ebbmath}{OML}{ccm}{m}{it}
\SetMathAlphabet{\ebbmath}{bold}{U}{eur}{b}{n}

\newcommand{\htwo}{H$_2$}
\newcommand{\rv}{$R_V$}
\newcommand{\ebv}{$E_{B-V}$}
\newcommand{\lya}{Ly-$\alpha$}
\newcommand{\Qa}{\ensuremath{Q_\mathrm{abs}}}
\newcommand{\Qe}{\ensuremath{Q_\mathrm{ext}}}
\newcommand{\Qs}{\ensuremath{Q_\mathrm{scatt}}}

\shortauthors{BURGH, McCANDLISS \& FELDMAN}
\shorttitle{FAR-UV DUST SCATTERING IN NGC 2023}

\begin{document}

\title{\large Rocket Observations of Far-Ultraviolet Dust Scattering in NGC 2023}

\author{Eric B. Burgh\altaffilmark{1}, Stephan R. McCandliss, Paul D. Feldman}
\affil{Department of Physics and Astronomy, The Johns Hopkins
       	University, Baltimore, MD 21218}
\altaffiltext{1}{Current Address: Space Astronomy Laboratory, University of
Wisconsin at Madison, 1150 University Avenue, Madison, WI 53706}

\received{}
\revised{}
\accepted{}

\ccc{}
\cpright{AAS}{}

\begin{abstract} 
The reflection nebula NGC~2023 was observed by a rocket-borne long-slit
imaging spectrograph in the 900~--~1400~\AA\ bandpass on 2000 February 11.  A
spectrum of the star, as well as that of the nebular scattered light, was
recorded.  Through the use of a Monte Carlo modeling process, the
scattering properties of the dust were derived.  The albedo is low,
0.2~--~0.4, and decreasing toward shorter wavelengths, while the phase
function asymmetry parameter is consistent with highly
forward-scattering grains, $g\sim0.85$.  The decrease in albedo, while
the optical depth increases to shorter wavelengths, implies that the
far-UV rise in the extinction curve is due to an increase in absorption
efficiency.
\end{abstract}


\section{Introduction}
The far-ultraviolet absorption and scattering properties of interstellar 
dust are not well determined, yet they are responsible for the regulation
of the penetration of far-UV light into interstellar clouds.  This energetic
radiation can ionize atomic species and photodissociate molecules, and thus 
plays an important role in the ionization, chemical, and thermal balances
of the interstellar medium.  The extinction curve along lines of sight through
the ISM is a measure of the combined effects of absorption and scattering by 
dust, and several studies indicate a relationship between the morphology of 
the far-UV extinction curve and diagnostics of the conditions within 
interstellar clouds \citep{Joseph89,Jenniskens92, Burgh00}.

However, it is important to note that the actual attenuation of radiation in a
cloud depends strongly on the far-UV scattering properties of the grains
responsible for the extinction, even when the extinction rises sharply in the
far-UV \citep{Flannery80}.  This is especially true for forward scattering
grains.  These will remove photons from line of sight observations toward
stars, producing extinction, but may not appreciably affect the radiation 
field within clouds, as the photons will be only slightly deflected from 
their original direction.
Thus, it is necessary to understand the breakdown of the extinction curve
into absorption and scattering efficiencies separately, as well as the
detailed properties of the dust itself, such as the albedo and the degree
of forward scattering.

Reflection nebulae provide useful sites for studying the scattering
properties of interstellar dust grains.  In many cases the illumination
is due to a single star and the geometrical distribution of the dust can
be constrained, or at least closely approximated. As the illuminating
source is contained very close to or within the nebula, its spectrum can
often be measured simultaneously with the nebular scattered light by the
same instrument.  This allows for the direct comparison of the source to
the scattered light independent of the instrument calibration.  The dust
scattering properties may then be derived through a comparison of the
observations to an appropriate model. 

A few bright nebulae have been studied in the ultraviolet,
including the Merope nebula \citep{Andriesse77,Witt85}, NGC~7023
\citep{Witt82,Witt92,Witt93,Murthy93}, the Scorpius OB association
\citep{Gordon94}, and IC~435 \citep{Calzetti95}.  This work reports on
the observations of the reflection nebula NGC~2023 with a far-UV
(900--1400~\AA), rocket-borne long-slit imaging spectrographic
telescope.

NGC~2023 is a bright reflection nebula located $\sim23\arcmin$ SE of
$\zeta$ Ori.  The illuminating star HD~37903 is of B1.5V spectral type
and is embedded in the L1630 molecular cloud at a distance of
450--500 pc \citep{deBoer83}.  It is one of the brightest reflection
nebulae in the sky and well-studied at many wavelengths including
[\ion{C}{2}] 158 $\mu$m \citep{Howe91,Jaffe94}, carbon recombination
lines \citep{Knapp75,Wyrowski00}, far-infrared dust emission
\citep{Harvey80,Mookerjea00}, and infrared \htwo\ emission
\citep{Gatley87,Field98,McCartney99,Martini99}.

A photometric study of the nebular surface brightness was performed by
\citet[WSK hereafter]{Witt84a} at visible wavelengths.  In their work, a
radiative transfer model that incorporated multiple scattering was used to derive the dust scattering
properties from photometric observations of the nebular surface
brightness profiles in the $uvby$ and $BVRI$ bandpasses ranging from
3500 and 10000~\AA.  In this work, this method is extended to
far-ultraviolet wavelengths; because the optical depths increase to
shorter wavelengths, it is very important to use a similar detailed radiative
transfer model that accounts for multiple scattering.

\section{Instrument and Observations}
\subsection{Instrument}

The telescope flown was a modified version of the Faint Object Telescope
(FOT) developed by the Johns Hopkins Sounding Rocket Group
\citep{Hartig80}.  The FOT comprises a 40~cm diameter, f/15.5 Dall-Kirkham
telescope with a long-slit imaging spectrograph.  To maximize the
efficiency in the far-UV bandpass, the optics were overcoated with ion
beam deposited SiC \citep{KeskiKuha88}.  At the focus of the telescope is
a mirror, which acts as the field stop for a slitjaw viewing camera and
has cut into it a long, narrow (200\arcsec x 12\arcsec) slit, defining
the entrance aperture to the spectrograph.

The spectrograph employs a 400 mm diameter Rowland Circle mounting with
$\alpha=8\fdg6$ and $\beta=0\arcdeg$ \citep{McCandliss94}.  It contains
a holographically-ruled, astigmatism-corrected diffraction grating used
in $-1$ order.  The spectrograph is sealed with a motorized gatevalve,
which operates during flight.  The detector is a Z-stack of microchannel
plates with a KBr photocathode and a delay-line anode for readout
\citep{Siegmund93}.  Additionally, an onboard calibration lamp
\citep{McCandliss00} was flown to provide an in-flight wavelength
calibration.

\subsection{Observations}

The sounding rocket experiment was launched aboard a Black Brant IX
sounding rocket from White Sands Missile Range, New Mexico ($106\fdg3$
W, $32\fdg4$ N) on 2000 February 11, at 20:27 MST.  The entire
flight time was approximately 15 minutes, with 400 seconds (from T+100
to T+500) spent with the detector high voltage on.  NGC~2023 was
acquired at T+175 and pointing stability was established through the use
of a wide field-of-view ($\pm1\arcdeg$) startracker, which maintained
``lock'' on $\zeta$ Ori during the observations.

The central star of NGC~2023, HD~37903, was observed for 125 seconds. 
However, the pointing was stable for only 98.5 seconds, with the star well 
situated in the slit and a steady count rate.  Only these data are considered 
here and are called the ``on-star'' pointing.
A far-UV spectrum of the star was recorded, as well as the nebular
scattered light to a distance of $\pm60\arcsec$ along the slit, which 
was oriented in the north-south direction.  The remainder of the
flight was divided into three offset pointings to explore the scattered
light distribution and to search for fluorescent emission from molecular
hydrogen (\htwo).

\section{Data}
The data from the rocket experiment were telemetered to the ground in the 
form of a bit stream.  For every detected photon, as well as calibration 
lamp photons, dark counts, and possible detector hot spots, $X$, $Y$, and
$T$ information was encoded in the stream.  $X$ and $Y$ correspond to the 
position on the detector of the event and $T$ the time of the event during
flight.  These data may be binned spatially and/or temporally.

\begin{figure}
\plotone{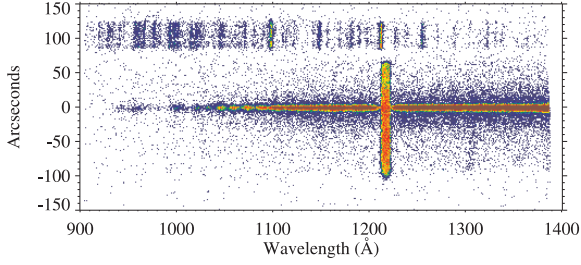}
\caption{Long-slit spectral image of NGC~2023.  The long axis of
the slit was oriented with north at the top and south at the bottom.
The spectrum at the top is that of the onboard calibration lamp.  Red
indicates the highest count rate regions.
\label{fig1}}
\end{figure}

A spectral image was created, shown in Figure \ref{fig1}, by compressing
the entire flight's data into an $X$-$Y$ image.  The image has been
corrected for several detector effects, as well as for spacecraft jitter
during the ``on-star'' portion.  Present in the image are the continuum
flux from HD~37903 (the horizontal stripe) and the nebular scatter
surrounding it.  The strong vertical line at 1216~\AA\ is the \ion{H}{1}
Ly-$\alpha$ airglow emission. Airglow lines at 1026~\AA\ and 1304~\AA\
are also present.  At the top of the image is a spectrum of atomic,
ionic, and molecular nitrogen provided by the onboard calibration lamp.
As the light from the lamp did not enter the spectrograph through the
entrance slit, there is a slight zero-point offset in the dispersion
direction.

\begin{figure}
\plotone{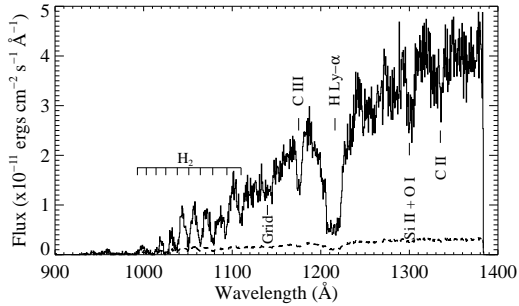}
\caption{Flux calibrated spectrum
of HD~37903.  The error in the flux is shown as the dashed line. Strong
stellar and interstellar absorptions are identified.  The feature marked
``grid'' is an instrumental artifact.
\label{fig2}}
\end{figure}

The calibrated stellar flux, computed from the central 10 pixels of the
detector, is shown in Figure \ref{fig2}.  The flux is consistent with
measurements made of the star by the Hopkins Ultraviolet Telescope (HUT)
\citep{Buss94}.  Clearly seen in the spectrum are several interstellar and
stellar absorption features, including those of \ion{H}{1},
\ion{C}{2}, \ion{C}{3}, \ion{O}{1}, \ion{Si}{2}, and the H$_2$ Lyman and
Werner bands.  Also, the strong effect of dust extinction is apparent,
since the intrinsic spectral energy distribution of a B1.5V star 
rises to shorter wavelengths.  The feature 
marked ``grid'' is an
instrumental artifact due to the vignetting of light by the QE
enhancement grid on the detector.

\begin{figure}
\plotone{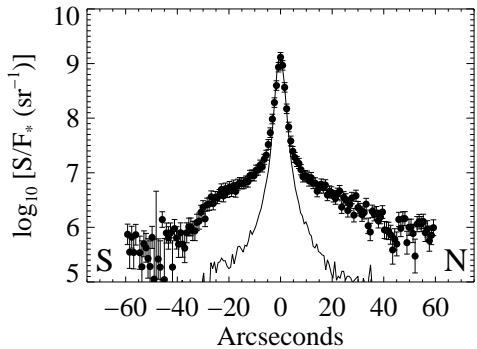}
\caption{Spatial profile of
nebular scattered light.  Filled circles with error bars are the data,
the solid line is the instrumental line spread function (LSF).  The 
profile is centered on HD~37903 with positive offsets to the north.
\label{fig3}}
\end{figure}

Due to the two-dimensional nature of the long-slit imaging spectrograph,
a spatial profile of the nebular scattered light can also be extracted from
the data.  The data were compressed in the spectral dimension, excluding the 
region around Ly-$\alpha$, binned up spatially along the slit as a surface
brightness, assuming that the scattered light fills the slit, and referenced
to the flux of the star.  This produced a slit profile of the ratio of the 
nebular surface brightness to stellar flux ($S/F_*$) (see Figure
\ref{fig3}), which is independent of the absolute detection efficiency of
the spectrograph. Overplotted is the instrumental line spread function (LSF)
as determined from laboratory measurements using a windowless vacuum
ultraviolet collimator \citep{Burgh01}.
The inner region is dominated by light from the star, but outside of 
$\pm10$\arcsec\ there is negligible contribution from the LSF and the nebular 
flux can clearly be observed, falling off exponentially with distance from the 
central star.  To the south of the central star, a dust lane appears to 
intersect the nebulosity, causing a depression in the flux centered at
$\sim50$\arcsec\ south.

\section{Extinction Curve and \rv}
 \label{extinctioncurve}

The derivation of dust scattering properties in reflection nebulae is
strongly model dependent.  Relevant parameters include the physical depth of
embedment of the illuminating star in the dust cloud and the wavelength
dependence of the dust opacity.  A depth of 1~pc was assumed by 
WSK.  This is consistent with the size of the nebula as
determined from the emission measure of carbon recombination lines
\citep{Knapp75}.  At the distance of the L1630 cloud, this corresponds to
$\sim450$\arcsec.
The wavelength dependence of the dust optical depth is derived from the
extinction curve; however, it is very important to determine what fraction
of the extinction along the line of sight to the illuminating star is
associated with the nebula itself and what fraction is due to foreground
material.  The optical extinction through the L1630 cloud is estimated
at $A_V\sim40$~mag \citep{Harvey80}; because the visible extinction
toward HD~37903 is $A_V\sim1.4$~mag, it is clear that the star is
located on the near side of this dense cloud.  From far-infrared (FIR)
measurements, \cite{Harvey80} suggest that the extinction from
FIR-emitting dust should account for a significant fraction of the total
extinction.  Since this dust emission is due to the processing of UV
photons, it is likely that most of the extinction along the line of
sight is associated with NGC~2023.

To determine the extinction curve, the ``pair method'' was used, in which 
the spectrum of an unreddened star of the same spectral type is compared to
that of HD~37903.
Following \cite{deBoer83}, HD~37744 was used as the comparison star for
HD~37903.  This star, only $\sim$~37\arcmin\ away from HD~37903, lies to
the west of the L1630 cloud and exhibits only a mild amount of
reddening (\ebv\ = 0.05) \citep{Hardie64,Crawford71}.  Its spectroscopic
parallax suggests that it lies only $\sim50$~pc more distant than HD~37903
\citep{deBoer83}, and as such samples essentially the same foreground
extinction.  The difference in reddening is thus due to the dust in the
L1630 cloud that lies between the front surface of the cloud and the
embedded star, and the extinction curve derived from comparing these two
stars should be indicative of the extinction properties intrinsic to the
nebular dust, canceling out the effects of the foreground dust along the
Orion line of sight.

\begin{figure}
\plotone{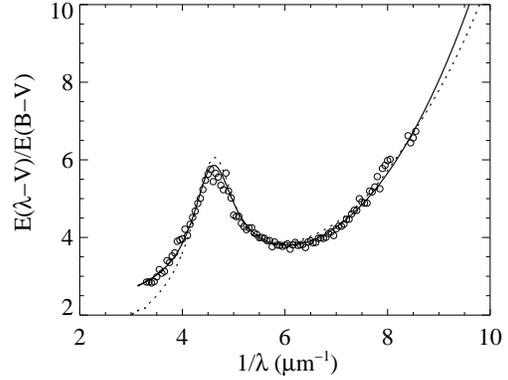}
\caption{Derived extinction curve for HD~37903 from
\textit{IUE} data.  The solid line is a fit following the
parameterization of \citet{FM90}.  The dashed line is the extinction curve
following the formula of \cite{CCM89} for $R_V=3.83$.
\label{fig4}}
\end{figure}

Low-dispersion data of both HD~37903 and HD~37744 were obtained from the
\textit{International Ultraviolet Explorer}
(\textit{IUE}) archive.  For each star, the short- and the long-wavelength
data were joined to a common wavelength grid ranging from 1150 to
3350~\AA.  The extinction curve, shown in Figure \ref{fig4}, was derived from the ratio of the
stellar fluxes, normalized to the difference in \ebv, and fitted with
the parameterization of \cite{FM90}.  The extinction curve,
$k(x)\equiv E_{\lambda-V}/E_{B-V}$ with $x=\lambda^{-1}$, is converted to
optical depth through the use of the ratio of total-to-selective
extinction, \rv\ [$\equiv A_V/E_{B-V}$], following $\tau=(k(x)+R_V) E_{B-V} /
1.086$.

The value for \rv\ was reassessed for this work.  The extrapolation of the
extinction curve to infinite wavelength, or zero frequency, at which
absolute extinction ($A_\lambda$) would be zero, provides for the
calculation of $R_V$.  The value of $R_V$ thus determined depends
strongly on the functional form of the extinction toward zero frequency.
It is useful to have extinction data at low frequency and near-infrared
(NIR) photometric observations are typically used.

It has been found that the shape of the NIR extinction curve is very
uniform along almost all lines of sight 
\citep{Rieke85,Whittet88,CCM89,Martin90b}, taking the form of a power law
($A_\lambda \propto \lambda^{-\alpha}$) between $\sim1~\mu$m and 5~$\mu$m.
At longer wavelengths, emission from dust can contaminate photometric
observations and complicate the determination of extinction.
\citet{Martin90b} expand this power law in terms of the extinction curve and \rv\ in the following form:
\begin{equation}
\frac{E_{\lambda-V}}{E_{B-V}}=e\lambda^{-\alpha}-R_V.
\end{equation}

\begin{figure}
\plotone{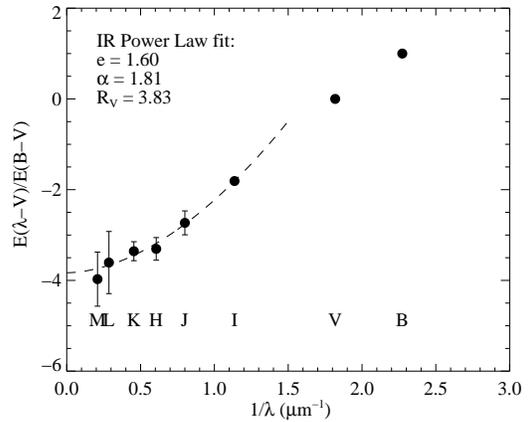}
\caption{Infrared color excesses with extinction fit
overplotted.  Shown are the best fit parameters following the offset
power law of \citet{Martin90b}.
\label{fig5}}
\end{figure}

For HD~37903, the NIR photometric data from the literature, listed in Table
\ref{table1}, were averaged together and the colors $\lambda-V$
determined.  The photometry for $V$ and \bv\ are taken from \citet{Hardie64}
($V$=7.84, \bv=0.10), \citet{Lee68} ($V$=7.83, \bv=0.11), and
\citet{Racine68} ($V$=7.82, \bv=0.09).  These were averaged together
to produce a $V$ magnitude and \bv\ color. By comparing the NIR colors
to the intrinsic colors for a B1.5V star
\citep{Koornneef83}, the color excesses $E_{\lambda-V}$ were computed.
These were then normalized by \ebv\ and a least squares fit was
performed with the \citet{Martin90b} power law described above (see
Figure \ref{fig5}).  This method produced a determination of
$\alpha=1.8\pm0.9$ and $R_V=3.83\pm0.40$.  The large error in these
values is mainly due to the uncertainty in the $L$ and $M$ band
photometry.

These results compare well with previous determinations of $\alpha$ and
\rv.  \citet{Whittet88}, from an analysis of an extensive data set
obtained from the available literature at the time, found an
$\alpha=1.70\pm.08$. \citet{CCM89} used the \citet{Rieke85} curve, which
has an $\alpha\simeq 1.6$ and \citet{Landini84}, through observations of
the hydrogen recombination lines in \ion{H}{2} regions, found values for
$\alpha$ and \rv\ of $1.85\pm0.05$ and 4.11 respectively.
\rv\ can also be
computed from color excesses following $R_V=1.1 E_{V-K}/E_{B-V}$, from the
standard van der Hulst theoretical curve, and gives a value of $R_V=3.7$.
These are comparable to the value of \rv\ based on the above calculation
within the error introduced by the uncertainty in the NIR photometric
measurements.

\section{Monte Carlo Model Fitting}
\label{montecarlo}

An analytic solution for the surface brightness of a reflection nebula
can be computed in the case of single scattering.  However, this would
only be accurate for optically thin nebulae.
\citet{Roark74} have shown, through the use of Monte Carlo
methods, that multiple scattering can have a significant effect on the
observed surface brightness distribution.  The intrinsic random nature
and statistical approach to the scattering process has allowed for
radiative transfer models of dusty nebulae to be constructed, including
some with arbitrary dust density and source distributions 
\citep{Witt77,Yusef84,Wood99,Gordon01}.

To determine the dust scattering properties in NGC~2023, a Monte Carlo
dust radiative transfer model was developed following the algorithm
described in full detail by \citet{Gordon01}.  To simplify the modeling 
process and reduce the computational demands, the nebula was modeled as a 
single star embedded in a spherical dust cloud of constant density and finite 
extent.  

Using this Monte Carlo process, a model nebula can be created for a
given radial optical depth, angular size of the nebula, and the dust's albedo,
$a$, 
and phase function asymmetry parameter, $g$, assuming the phase function of 
\citet{Henyey41}.  The first two are set and the dust properties $a$ and $g$
are varied until agreement with the observations are maximized.  The 
following section discusses the results of fitting models to the data and the 
resultant derivation of the dust scattering properties.

\subsection{Radial Profile Fitting}

The data were binned spatially to produce a slit profile, compressing the
entire wavelength range into a one-dimensional profile, excluding a
10~\AA\ band (one slit width) around the \lya\ geocoronal airglow line
at 1216~\AA.  The error for each bin was computed using Poisson
statistics, i.e. $\sigma_\mathrm{bin}=\sqrt{N}$, with $N$ the number
of counts in the spatial bin.  The inner $\pm$5\farcs5 of the profile is
dominated by the star and we take the total counts in this region to
correspond to the stellar flux (F$_*$).

The flat-field corrected ratio of the nebular surface brightness to stellar 
flux, $S/F_*$, for the ``on-star'' exposure is shown with
logarithmic scaling in Figure \ref{fig3}.  Positive offsets are
to the north. Outside of the inner region, where the instrumental LSF
causes the stellar flux to dominate, the nebular brightness falls off
exponentially by about an order of magnitude.  Only the region inside of
$\pm60$\arcsec\ contained data with a count rate above the dark rate.
To the south, a dust lane cuts in front of the nebulosity at around
$-50$\arcsec.

\begin{figure}
\plotone{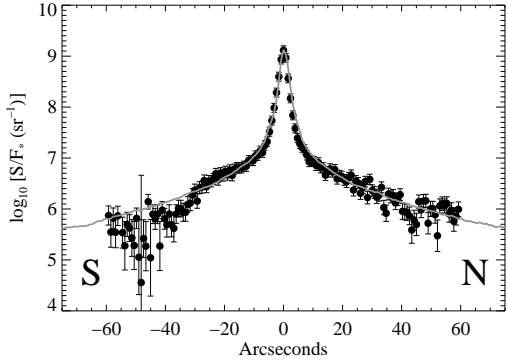}
\caption{Slit profile of nebular surface brightness to
stellar flux ratio for the ``on-star'' position.  These data encompass
the entire detector bandpass, excluding \lya.  Overplotted is the
model corresponding to the best fit $a$ and $g$.  The profile is centered on
HD~37903 with positive offsets to the north.
\label{fig6}}
\end{figure}

A grid of models for various values of $a$ and $g$ corresponding
to the far-UV optical depth was generated.  The optical depth used was a
stellar flux intensity weighted average of the $\tau(\lambda)$ derived
from the extinction curve in Section \ref{extinctioncurve}.
For each
model in the grid, the $\chi^2$ statistic was computed by comparing the
model to the data following
\begin{equation}
\chi^2 = \sum_i \left(\frac{y_i - y(x_i;\tau,a,g)}{\sigma_i}\right)^2 .
\end{equation}
In this definition, $y_i$ and $\sigma_i$ are the data and their
respective errors for the $i$th spatial bin.  The term $y(x_i;\tau,a,g)$ 
is the model prediction
for $S/F_*$ as a function of radius for the given optical depth and the
input values for $a$ and $g$.  The sum is over the region
$r=10-60$\arcsec\ on both the north and the south sides of the central
star.  The instrumental LSF was added in to the model fits; however, its
presence had no effect on the fitted parameters as the contribution from
the star due to the LSF is negligible outside of 10\arcsec.  The best
fit parameters were determined from the model that gave the minimum
value of $\chi^2$.  The best fit parameters for the full experiment
bandpass are listed at the top of Table \ref{table2} and the model is
shown overplotted on the data in Figure \ref{fig6}.  Additionally, as
the profile showed some asymmetry, the north and south sides of the star
were fitted separately.  These produced different fit results; however, that
may be more likely due to a variation in the dust distribution than
actual differences in the scattering characteristics.

To obtain the wavelength dependence of the dust scattering properties,
the data were divided into several wavelength regions.  Six wavelength
regions were chosen across the full bandpass, and the above procedure
was used to fit these data, with a few exceptions.  The spatial range
was limited to less than 60\arcsec.  This is due to the fact that as
fewer data were used the contrast between the nebular scattered light
and the background was worse and the outer limit of the range was
decreased to avoid fitting the background. The maximum fitting radius was 
chosen for each bandpass so that the signal-to-noise ratio ($S/N$) of the 
spatial bins was always greater than one.  The widths
of the bandpasses were chosen so that this radius was greater than
30\arcsec.  The wavelength ranges and maximum radii are listed in Table
\ref{table2}.
For the shorter wavelength bandpasses, in which there are considerably
less data due to the line-of-sight extinction, a coarser spatial binning
was used.  This provides more counts per spatial bin, which results in a
higher $S/N$ per bin, and thus a larger radius before the $S/N$ dropped to
one.  If the maximum radius were too close to the star, the small number
of data points would limit the constraint on $g$.  The best fit dust
parameters are listed in Table \ref{table2}.

\begin{figure}
\plotone{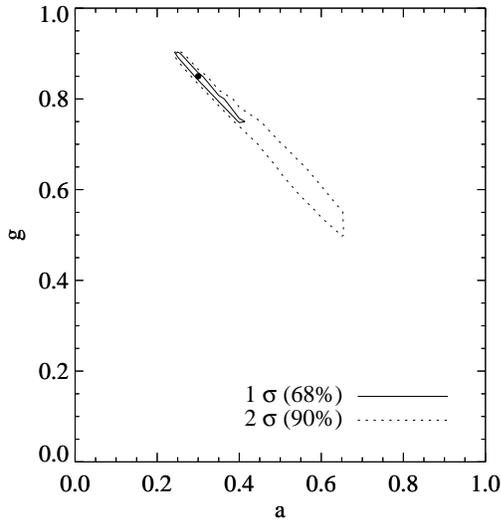}
\caption{Confidence intervals from the $\chi^2$ contours for the model fit 
of the nebular brightness radial profile for the 1240--1295~\AA\ bandpass.  
The black dot indicates the best fit parameter values, corresponding to the 
minimum $\chi^2$.  There is a degeneracy between the 
two parameters and the confidence tends toward lower $g$ and higher $a$.
The confidence intervals for the other bandpass fits are essentially of
similar shape.
\label{fig7}}
\end{figure}

Confidence intervals were based on the $\chi^2$ contours.  Figure
\ref{fig7} shows the  $\chi^2$ contours for the
1240--1295~\AA\ bandpass.  The $\chi^2$ fitting produces asymmetric
error bars, with most of the confidence tending toward lower $g$ and
higher $a$.  Thus, it is apparent that there is a strong coupling
between $a$ and $g$.  Any random or systematic effects that would change the 
value of $g$ could have a corresponding change in $a$.

\begin{figure}
\plotone{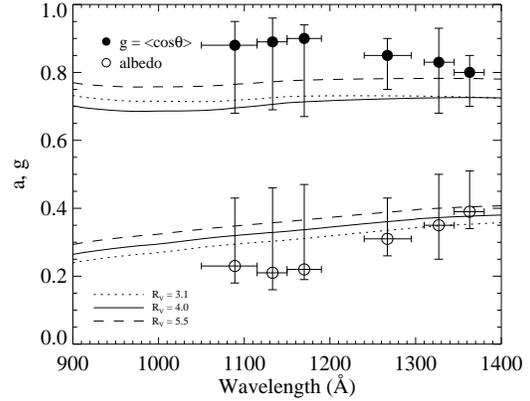}
\caption{Derived values of $a$ and $g$ for the profile fits
listed in Table \ref{table2}.  Theoretical predictions for various
values of \rv\ from \citet{Weingartner01} are overplotted.
\label{fig8}}
\end{figure}

The derived dust parameters from Table \ref{table2} are plotted in
Figure \ref{fig8}.  For comparison, the predicted values from
\citet{Weingartner01} for values of $R_V$ of 3.1, 4.0, and 5.5 are
overplotted.  We find $g$ increases mildly across the bandpass, while
$a$ falls toward shorter wavelengths.

\subsection{Spectral Fitting}

To achieve a better $S/N$, the data can be binned up
spatially and the ratio of the nebular surface brightness to stellar
flux shown as a function of wavelength rather than slit position.
However, this removes the ability to fit $g$ in an independent manner,
so for this procedure a constant value of $g$ was assumed, as supported 
by the results of the previous section.  Then, as a
function of wavelength, the $S/F_*$ can be used to derive the albedo directly
through the model fitting.

\begin{figure}
\plotone{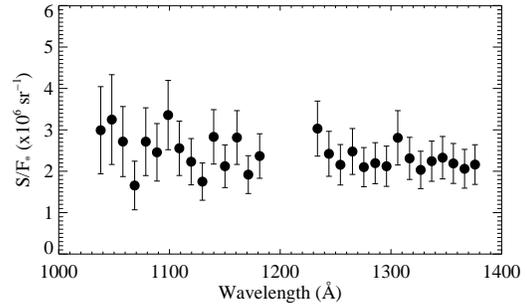}
\caption{Spectrally binned $S/F_*$.  The region around
\lya\ is excluded.
\label{fig9}}
\end{figure}

The data are binned in the spectral dimension,
compressing the spatial range from 10~--~60\arcsec\ into one bin for each
wavelength range.  Because the nebular scattered light is a diffuse
source and is expected to fill the slit, the data were binned by one
slit width, or 10~\AA\ across the full wavelength region, excluding the
\lya\ region, to 1030~\AA.  At shorter wavelengths, the detected count
rate for the nebular scattered light is consistent with the background.
The stellar spectrum was also smoothed by one slit width before the
ratio was determined.  The derived $S/F_*$ is shown in Figure \ref{fig9}.
It is virtually constant at all wavelengths, despite a rising optical
depth toward shorter wavelengths.

\begin{figure}
\plotone{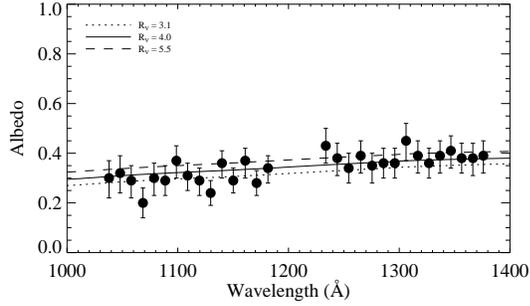}
\caption{Derived wavelength dependence of albedo from model
fitting of the spectrally binned data.  The prediction for $R_V$ values
of 3.1, 4.0, and 5.5 from \citet{Weingartner01} are overplotted.
\label{fig10}}
\end{figure}

The data are compared to a grid of models with different albedos for an
assumed value of $g=0.85$ across the bandpass.  Since the $g$ from
the radial profile fitting was fairly constant, the average value is
used in the fitting here.  The best fit albedos are shown in Figure
\ref{fig10}.  There is a decrease in albedo as the wavelength 
decreases.  This is consistent with the results derived from the profile
fitting above (see Figure \ref{fig8}).  Overplotted are the
predictions from \citet{Weingartner01} for a few values of \rv.  The
derived albedos agree very well with the predictions.

\begin{figure}
\plotone{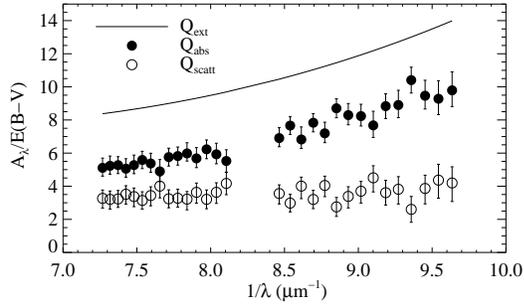}
\caption{Relative values of the extinction, absorption, and
scattering efficiencies derived from the best fit albedos. The dust
extinction efficiency, $\Qe$, is proportional to the extinction curve
for HD~37903.
\label{fig11}} 
\end{figure}

The relative efficiencies for both scattering and absorption, $\Qs$ and
$\Qa$, can be derived from the albedo, since $a = \Qs/\Qe$.  Therefore,
$\Qs=a\Qe$ and $\Qa=(1-a)\Qe$.  The extinction efficiency, $\Qe$
($=\Qa+\Qs$), is proportional to the extinction curve, and thus to
$\tau$.  The derived relative efficiencies are shown in Figure
\ref{fig11}, along with the extinction curve, in units of
$A_V/$\ebv.  It is apparent that the rise in the far-UV extinction curve
of NGC~2023 is due to an increase in absorption, whereas the scattering 
efficiency is nearly constant.

\subsection{Systematics}

As mentioned earlier, the derivation of the dust parameters $a$ and $g$
is sensitive to the modeling process.  There are several sources of
systematic error that can affect the best fit parameters.  This section
discusses some of these sources and describes the effects they have on
the derived dust properties.

In the data preparation process, it was assumed that all of the detected
photons within $\pm$5\farcs5 of the peak of the radial profile are
stellar in origin.  However, a contribution from scattered light at
small angles, which would be included with the stellar light, would result 
in an underestimation of $S/F_*$ and hence the derived albedo.
This effect would become increasingly important as all the parameters,
$\tau$, $a$, and $g$, increase.  As the optical depth increases, the
contrast between the escaping stellar light, F$_*$, and the nebular
scattered light would decrease due to the increased probability of
interaction.  An increase in albedo would make the nebular scattered
light brighter, and higher values of $g$ increase the chance of
scattering into the line of sight at small angles from the star.

Assuming that the constant density of the model can be continued to the
position of the star, the Monte Carlo model was used to estimate the
number of nebular photons near the center that would be mixed in with
the stellar light.  For $\tau=2-3$, $a=0.2-0.4$, and $g=0.8-0.9$, the
fraction of nebular contribution to the stellar flux ranges from
5~--~20\%.  However, based on \ion{C}{2} emission, \citet{Knapp75}
conclude that there may be a dust-free \ion{H}{2} region surrounding
HD~37903 with a radius as large as $\sim25$\arcsec, and \citet{Harvey80}
suggest that the inner 0.04~pc (18\arcsec) is clear of dust.  A ``hole''
in the dust distribution would reduce the contribution from scattered
light in the vicinity of the star, and the modeled contamination is thus
an upper limit.  Decreasing the stellar flux in the model fits as
described above resulted in an increase in the fitted albedo by an
amount proportional to the flux reduction, but did not affect $g$;
i.e., a decrease in flux by 10\%\ corresponded to an increase in the
derived value of $a$ by 10\%.

A spherical geometry was assumed in the modeling process.  For the L1630
cloud, a plane-parallel slab model may be more appropriate.  However, if
the depth of embedment, and thus the radius of the sphere in the model,
is large relative to the extent being observed as projected onto the
sky, the difference between the sphere and slab models is negligible.
We used the value from WSK of 1~pc for the radius, and therefore the
model fits were performed on the inner 0.5--1\farcm0 of an assumed
$\sim7$\arcmin\ nebula.  If the depth of embedment, and hence the radius
of the spherical model, is of lower value, then the fitted $g$ values
decrease.  Also, the correlation of $a$ and $g$, as apparent from the $\chi^2$ 
confidence intervals shown in Figure \ref{fig7}, is weaker for smaller 
depth.

At too small a radius relative to the observed extent, the assumption of
spherical symmetry is no longer sufficient to describe the geometry.  In
this case, it is possible that the observed nebular surface brightness
distribution can be explained by a slab geometry of shallow depth but
large horizontal extent.  There do exist degeneracies between the two
geometries, such that a spherical cloud of given $a$ and $g$ can appear
identical to a slab with the star embedded less deep and lower values of
$g$ but observed out to radii larger than the actual depth of
embedment.  The value of 1~pc for the radius produces fits to the visible
wavelength surface brightness with reasonable values for $a$ and $g$
(WSK).

The optical depth used in the modeling is dependent on the far-UV
extinction curve.  The primary systematic error affecting
$\tau_\mathrm{UV}$ is from the propagation of the $L$ and $M$ band
photometric errors in the determination of \rv.  Although these were
large, the resultant uncertainty in \rv\ does not have a large effect on
the optical depth in the UV and the fitted parameters are modified only
slightly ($\lesssim10$\%).  If less of the extinction were associated
with the nebula itself, the albedo would increase; however, not more
than half can be foreground or the far-UV albedo would be driven to the
maximum value of 1.0.  There is also an implicit assumption that,
for the lack of a far-UV spectrum of HD~37744, the UV extinction curve
as derived from \textit{IUE} can be extrapolated to shorter wavelengths than
1150~\AA\ following the parameterization of \citet{FM90}.

As already mentioned, a constant dust density
distribution was assumed in the modeling process.  This greatly reduced
the computation time, because the distance each photon traveled was simply 
linearly proportional to the optical depth along all directions through the 
model dust cloud.  WSK, on the other hand, assumed a weak radial falloff in
the density distribution.  Such a radial density dependence could
account for some of the observed falloff in nebular surface brightness
and reduce the value of $g$ necessary to fit the slit profiles.
Although there is the possibility of large scale density variations, the
nebula appears to be fairly smooth in both the blue, as observed by the
digitized sky survey, and the far-UV, as shown in the slit profile in
Figure \ref{fig3}.

It should be noted that the predicted dust properties, such as those of
\citet{Weingartner01}, to which the results have been compared, are
based on Mie scattering theory, which solves for the exact
electromagnetic fields in a spherical or cylindrical dust grain.
However, it is clear from polarization studies that dust grains in the
interstellar medium are not spherical.  The more probable situation is
that dust grains collide and either stick together or shatter, producing
grains of a more random distribution of shapes
\citep[e.g.]{Mathis89,Bazell90,Fogel98}.  \citet{Savage79} point out
that the Henyey-Greenstein phase function is an analytic function that
approximates the results from Mie theory and may not represent the true
scattering phase function.

\citet{Witt93} have pointed out that dust scattering is often
accompanied by fluorescent emission of \htwo.  It is therefore necessary
to account for any contribution from the \htwo\ far-UV fluorescent
cascade to the dust scattered continuum.  Fluorescently excited \htwo\
is indeed present as evidenced indirectly by the detection of absorption
lines from highly excited vibrational states \citep{Meyer01}.  More
direct evidence is offered by the \textit{IUE} spectrum, SWP~38082,
which exhibits the same characteristic pair of fluoresecent \htwo\
emission lines at 1577 and 1608 \AA\ as seen in the IC~63 (SWP~33898)
spectrum where \htwo\ fluorescence has been confirmed \citep{Witt89}.
The scattered continuum in NGC~2023 has an $S/F_*\sim10^{5.6}$ in the
wavelength region longward of 1700~\AA, and the \htwo\ emission peaks at
$\sim$~3 times this value.

However, the \textit{IUE} spectrum of NGC~2023 was acquired at a
position $\sim$~75\arcsec\ east-north-east of HD~37903, coincident with a
bright infrared emitting knot as revealed in high spatial resolution
images of the nebula.  These images, which isolate specific vibrational
quadrupolar transitions in the ground electronic state of \htwo,
generally show a clumpy distribution \citep{Field98,Takami00}.  In
particular, the \citet{Takami00} images shows the region to the north of
HD~37903 to be devoid of \htwo\ infrared emission within the extent of
the rocket experiment's slit.  To the south there are two lanes; one is
rather close, $\sim$~10~--~20\arcsec, to the central star and comparable
in brightness to the east-north-east knot, and the other (the southern
bar) is $\sim$~70\arcsec\ south of HD~37903 and brighter, with a void of
emission in between.  \citet{Martini99} have observed the infrared
emission spectroscopically in the region to the south of the star that
is spatially coincident with our slit.  They find that the brightest infrared
\htwo\ emission lies in a bar between 67\arcsec\ and 84\arcsec\ south
that is $\sim$~3.5 times stronger than the region closer to the star.
They also show that the infrared continuum emission is essentially negligible
in the southern bar and increases towards the star, becoming comparable
to the \htwo\ emission close in.

The southern bar of the nebula was observed within our slit; however,
the data here are consistent with the background, with a 2 $\sigma$
upper limit of $S/F_*=10^{5.5}$.  As this is the brightest region of
\htwo\ infrared emission along the slit, we conclude that the region
closer to the star, in which we detect $S/F_*=10^{6.5}-10^7$, has a
neglible contribution from far-UV \htwo\ fluorescent emission.  The
asymmetry between the profiles in the north and the south sides of the
star mentioned previously is thus more likely due to asymmetries in the
dust distribution than additional emission from \htwo.  This asymmetry
has only a small effect on the fitting of the modeled slit profile to
the data.

We note that the rocket flight included two offset pointings in the
north-eastern portion of the nebula, $\sim$45\arcsec\ from the star,
which do show an excess of emission from the best-fit scattered light model
by a factor of $\sim2-4$.  Although the spectral resolution of the
sounding rocket is ideal for investigating dust scattered light it is
insufficient for unambiguously detecting suspected far-UV fluorescent
emissions from \htwo.

\section{Discussion and Conclusions}
The results described above show that the dust in the reflection nebula
NGC~2023 is highly forward scattering and has a low albedo that
decreases with decreasing wavelength. These results are in good agreement
with the predictions of $a$ and $g$ based on the fitting of extinction
curves \citep{Li01,Weingartner01}.  The rise in the far-UV
extinction curve is due to an increase in the absorption
efficiency of the dust grains, whereas the scattering efficiency
appears constant.  This suggests that the scattering is being performed
by large grains, while a low-albedo small grain population may be the
cause of the short wavelength absorption.  

The approach taken here is very similar to that of previous work on
other reflection nebulae.  NGC~7023, for example, is a similar, but more
extincted, nebula to NGC~2023.  It has been observed extensively in the
ultraviolet, by \textit{IUE} \citep{Witt82}, the Ultraviolet Imaging
Telescope (UIT) \citep{Witt92} and the Hopkins Ultraviolet Telescope
(HUT) \citep{Murthy93} as part of the Astro-1 mission, and 
$Voyager$~2 \citep{Witt93}.

The initial study of NGC~7023 with \textit{IUE} suggested more isotropic
scattering in the UV than the visible, but the measurements with UIT
provided more detailed surface brightness profiles and \citet{Witt92}
conclude that the dust in that nebula is forward scattering
($g\sim0.70$).  The UIT measurements indicated a high albedo
($a\sim0.6$), similar to that in the visible, but the short wavelength
limit was 1400~\AA.  The HUT measurements extended into the windowless
ultraviolet bandpass covered by the sounding rocket experiment.
\citet{Murthy93}, using a similar Monte Carlo method, also found that
the albedo was high ($\sim0.6-0.7$) in the 1000--1400~\AA\ range, but
decreasing to shorter wavelengths.  They note that their derived albedo
would decrease if more of the dust can be associated with the nebula.
\citet{Witt93} show models for three different optical depth radii in
their fitting of the $Voyager$ data, resulting in dust albedos at
$\lambda=1000$~\AA\ of 0.45~--~0.26, with each case showing a clear
decline (about 25\%) when going from 1400 to 1000 \AA.

\citet{Calzetti95} observed the reflection nebula IC~435, which is very near
NGC~2023 in Orion.  Their data were taken with \textit{IUE} and were
intended to investigate the scattering properties near the 2200~\AA\
bump.  Also employing a Monte Carlo model, they derived the dust albedo
and phase function asymmetry parameter.  The results indicate a drop in
albedo at the bump, but rising albedo toward the short wavelength limit
of their observations.

It might be expected that, because this nebula appears to also be in the
L1630 cloud, the dust characteristics would be similar to those of
NGC~2023.  However, HD~38087, the central star of IC~435, appears to
have a lower than average far-UV rise, and
\citet{CCM89} find an \rv\ of 5.3.  \citet{Aiello88} list a value
of \rv\ = 4.86 for HD~38087, but it is clear that the extinction toward
this nebula is more consistent with a dense region where the dust grains
may be larger than average.  Also, as this star is of spectral type B5V, the 
dust environment around the star may be different than that of NGC~2023 if 
any processing of the dust grains occurs due to the high ultraviolet flux 
of the hotter HD~37903.  However, Calzetti et al. associate only
$\sim20$\%\ of the extinction along the line of sight to the nebula
itself. This would indicate that the foreground material, which would
therefore account for most of the extinction curve, has the dust properties
of a dense environment.  This is inconsistent with the calculation of
the extinction curve toward HD~37903 as described in Section
\ref{extinctioncurve}, which indicates little foreground extinction
toward the L1630 cloud.  Had they used a higher optical depth in their
Monte Carlo model, their derived albedos would be of lower value, more
in agreement with the results derived here.

Measurements of dust scattering properties through observations of the
diffuse galactic light (DGL) have significant scatter. Some recent
ultraviolet measurements have been made.  Observations with the
Berkeley UVX spectrometer \citep{Hurwitz91} indicate a low $a$, but also
a low $g$ in the 1400--1850~\AA\ bandpass. \citet{Witt97} found, with
the Far Ultraviolet Space Telescope (FAUST) on board the Atlas-1
mission, an albedo of 0.45 and $g=0.68$ in the same bandpass. 
\citet{Schiminovich01} obtained similar results from modeling the 
observations from a wide-field ultraviolet 
sounding rocket experiment, finding $a=0.45\pm0.05$ and $g=0.77\pm0.1$ at 
1740~\AA.  These slightly lower albedos for the diffuse ISM, relative to 
those found in the denser regions of NGC~7023 and IC~435, have been 
associated with a dust size distribution including more small grains than 
average.

There is a large spatial variability to Galactic interstellar
extinction, particularly in the ultraviolet.
\citet{Witt93} point out that if these variations are due to changes in
the contribution of various grain components to the total extinction
then it may be expected that nebula in differing environments may show
different far-UV dust albedos.

Though agreeing with the predictions of \citet{Weingartner01}, the
albedo of the dust in NGC~2023 appears to be more consistent with the
observations of lower albedo dust in the diffuse ISM as determined from
measurements of the DGL, even though it resides within a reflection
nebula with an \rv\ indicating a denser than average environment.  Some
of this discrepancy could be due to the assumption that the dust in the
nebula is homogeneous.  \citet{Witt96} have performed simulations of the
radiative transfer through two-phase clumpy dust of various filling
fractions.  Their models suggest that radiative transfer techniques
based on homogeneous models can underestimate the albedo when applied to
inhomogeneous nebulae.  This effect becomes more significant at higher
values of $g$ and $\tau$.  In their study of the infrared scattering
properties in NGC~2023, \citet{Sellgren92} suggest that the dust may be
clumpy in order to reconcile the $J$, $H$, and $K$ band surface
brightness measurements with the UV and visible measurements.

It is critical to determine the far-UV scattering properties of dust in
order to understand the depth dependent effects of the penetration of
far-UV photons into interstellar clouds.  In their study of the
photodissociation of carbon-bearing molecules, \citet[vDB
hereafter]{vDB89} show that varying the morphology of the extinction
curve can cause order of magnitude changes in the central photorates in
translucent clouds, in particular for CN and CO, which photodissociate
at wavelengths shorter than 1100~\AA.  These effects may have been observed
in the studies of \citet{Joseph89} and \citet{Burgh00}.  However, it
should be noted that the vDB study assumed the dust scattering
characteristics of Model 2 of \citet{Roberge81}, which has an $a=0.6$
and $g=0.5$.  These values are not consistent with the results found
here, nor with many of the other studies, nearly all of which find a
higher value of $g$ in the ultraviolet.  Had vDB used the dust
properties indicated in this study of NGC~2023, their photorates may
have been different.  The lower albedos indicated here would drop the
dissociation rates due to the stronger attenuation by the dust; however,
the more forward-scattering grains could possibly counteract that effect
by increasing the probability that a photon might get deeper into the
cloud before interacting with either a dust grain or molecule.

Although it is very important to determine the ultraviolet scattering
properties of dust, it is clear that there are several inherent difficulties in the modeling
of scattered light in reflection nebulae.  The most important being the
geometry and distribution of the dust relative to the illuminating
source.  Further studies of the far-infrared emission from dust, in
conjunction with these and other far-ultraviolet measurements of dust
scattered light, may help to constrain the geometry and thus limit the
introduction of systematic errors into the determination of the dust
scattering properties.

\acknowledgments The authors would like to extend their gratitude to
Russell Pelton for the vital role he plays in the continuing success of
the JHU Sounding Rocket Group.  We would like to acknowledge NASA's Wallops
Flight Facility and the NSROC contractor, as well as the support
personnel at the White Sands Missile Range and the Physical Sciences
Laboratory of NMSU.  We would also like to thank Brad Frey and Geoff
Brumfiel for technical assistance. Additionally, we are thankful for useful 
discussions with Daniela Calzetti and Richard Henry.  This work was 
supported by NASA grant NAG5-5122 to The Johns Hopkins University.

\clearpage

\begin{deluxetable}{ccccc}
\tablewidth{2.7in}
\tablecaption{Near-infrared photometry
of HD~37903.}
\tablehead{& & \multicolumn{3}{c}{Magnitude} \\
\colhead{Band} & \colhead{$\lambda$ ($\mu$m)} & \colhead{Ref 1} &
\colhead{Ref 2} & \colhead{Ref 3}}
\startdata
I	& 0.88 & 7.52 & &\\
J	& 1.25 & 7.48 &7.36&7.44\\
H	& 1.65 & & 7.28 & 7.35\\
K	& 2.20 & 7.41 & 7.28 & 7.32\\
L	& 3.50 & 7.00 & 7.44 & 7.36\\
M	& 4.80 & & & 7.20
\tablerefs{Ref 1 -- \citet{Lee68};
Ref 2 -- \citet{Whittet78};
Ref 3 -- \citet{DePoy90}}
\enddata
\label{table1}
\end{deluxetable}

\begin{deluxetable}{cc|c|c|cc}
\tablewidth{4.0in}
\tablecaption{Best fit $a$ and $g$ from radial profile fitting.}
\tablehead{\multicolumn{2}{c|}{Bandpass} & Max.$^\dagger$&&\multicolumn{2}{c}{Best-fit dust parameters}\\
$\lambda_\mathrm{min}$ (\AA) & $\lambda_\mathrm{max}$ (\AA) & radius
(\arcsec)& $\tau ^\ddagger$ &
$a$ & $g$}
\startdata
\phn990 & 1380  & 60 & 2.8 & 0.26 & 0.89\\
\hline
\phn990 & 1380  & 60 N & 2.8 & 0.27 & 0.87\\
\phn990 & 1380  & 40 S & 2.8 & 0.23 & 0.92\\
\hline
1050&1115& 33 & 3.6 & 0.23$^{+0.20}_{-0.05}$ & 0.88$^{+0.07}_{-0.20}$\\
1115&1150& 33 & 3.2 & 0.21$^{+0.25}_{-0.05}$ & 0.89$^{+0.07}_{-0.20}$\\
1150&1190& 35 & 3.0 & 0.22$^{+0.25}_{-0.03}$ & 0.90$^{+0.04}_{-0.23}$\\
1240&1295& 40 & 2.7 & 0.31$^{+0.12}_{-0.05}$ & 0.85$^{+0.05}_{-0.10}$\\
1310&1345& 40 & 2.5 & 0.35$^{+0.15}_{-0.10}$ & 0.83$^{+0.10}_{-0.15}$\\
1345&1380& 35 & 2.4 & 0.39$^{+0.12}_{-0.05}$ & 0.80$^{+0.05}_{-0.10}$
\enddata
\tablenotetext{\dagger}{Maximum radius of the fitting region}
\tablenotetext{\ddagger}{Intensity weighted optical depth for the
given bandpass}
\label{table2}
\end{deluxetable}

\clearpage

\end{document}